\documentstyle[preprint,eqsecnum,amssymb,aps]{revtex}
\tightenlines
\def\BZ{\mbox{$\Bbb Z$}} \def\BR{\mbox{$\Bbb R$}}
\begin{document}
\draft

\preprint{\hbox{\vbox{\tt \noindent BONN-TH-97-07~~\\ hep-th/9705113}}}

\date{May 14, 1997; Revised June 4, 1997, July 17, 1997}
  \title{\large {\bf A note on  M(atrix) theory in seven dimensions with eight
supercharges}}
\author{\it Suresh Govindarajan\cite{byline}}
\address{
Physikalisches Institut der Universit\"at Bonn\\
Nu\ss allee 12, D-53115 Bonn, Germany                \\
{\tt Email: suresh@avzw02.physik.uni-bonn.de}}
\maketitle
\begin{abstract}
We consider M(atrix) theory compactifications to seven dimensions with
eight unbroken supersymmetries.  We conjecture that both M(atrix)
theory on $K3$ and Heterotic M(atrix) theory on $T^3$ are described by
the same 5+1 dimensional theory with ${\cal N}=2$ supersymmetry broken
to ${\cal N}=1$ by the orbifold projection.  The
emergence of the extra dimension follows from a recent result of
Rozali ({\tt hep-th/9702136}). We show that the seven dimensional
duality between M-theory on $K3$ and Heterotic string theory on $T^3$
is realised in M(atrix) theory as the exchange of one of the
dimensions with this new dimension.
\end{abstract}

\pacs{PACS: 11.25.-w}

\section{Introduction}

M(atrix) theory has provided new insights into understanding
M-theory\cite{bfss} beyond its description using eleven dimensional
supergravity or as the strong coupling limit of IIA superstring
theory\cite{witten}.  Compactifications of M(atrix) theory preserving
16 supersymmetries have been useful in understanding U-dualities in
dimensions $d\geq7$\cite{comp,ttwo,tthree}.  For example, in $d=8$
part of the U-duality group is realised as the strong-weak coupling
duality of ${\cal N}=4$ Yang-Mills in four dimensions\cite{tthree}. In
the case of $d=7$, Rozali has shown that the full U-duality group
(=SL(5,\BZ)) is geometrically realised by including an extra dimension
(momentum modes in this dimension correspond to zero size instantons
on the four-torus)\cite{rozali}. In addition, string duality predicts
interesting properties for each of these Yang-Mills
theories\cite{fhrs}. The M(atrix) description of string theory leads
to second quantised string theory with the three point interaction
vertex being described by the crossing of eigenvalues in M(atrix)
theory\cite{string}.

Compactification of M(atrix) theory on $T^d$ leads to U(N) Yang-Mills
theory in $d+1$ dimensions on the dual torus\cite{bfss,comp}. 
However, Yang-Mills theory is non-renormalisable in dimensions larger than 
four. Thus the SYM prescription breaks down when $d$ is larger than
three. However there has been progress recently for M(atrix) theory on
$T^4$ as well as $T^5$\cite{brs,five}. For $T^4$, the proposed model is
a $5+1$ dimensional field theory with chiral ${\cal N}=2$ supersymmetry.

The simplest cases where lesser supersymmetry (eight unbroken
supersymmetries) occurs is for M(atrix) theory on ALE
spaces\cite{douglas}, M(atrix) theory on $K3$\cite{kthree} and
Heterotic M(atrix) theory\cite{dwall,het,horava}.  In seven
dimensions, it is well known that the latter two theories are dual to
each other -- this is the M-theoretic extension of the
well-established IIA - Heterotic string duality. Both these theories
can be obtained from different orbifolds of M(atrix) theory on
$T^4$. $K3$ in an orbifold limit is described by $(T^4/\BZ_2)$ while
the Heterotic case corresponds to the orbifold $(S^1/\BZ_2) \times
T^3$. The orbifold group acts on the Yang-Mills theory as:
$\widetilde{T}^4/\BZ_2$ and $\widetilde{S}^1 \times
(\widetilde{T}^3/\BZ_2)$ respectively for the K3 and Heterotic
theories.  Before orbifolding, this is the theory studied by
Rozali\cite{rozali} and one obtains the emergence of an extra
dimension. In both cases, the $\BZ_2$ orbifold group can be extended
to include the extra dimension.
By studying the action of the orbifold $\BZ_2$ on
the coordinates of the underlying 5+1 dimensional theory, we observe
that both compactifications can be identified on exchanging/relabeling
coordinates! U-duality permits us to treat the five coordinates
symmetrically and this leads us to conjecture that the two theories
are described by the {\it same} 5+1 dimensional theory\footnote{
The emergence of K3 in Heterotic Matrix theory has been observed
by P. Ho\v rava\cite{horava,dual}.}. The seven
dimensional duality mentioned earlier is shown to be a consequence of
this conjecture. By using this conjecture, we reproduce the
relationship between the Heterotic string tension and string coupling
constant to the $K3$ parameters as predicted by seven dimensional
duality.  As is clear from the arguments mentioned above, for the
conjecture to work we need to assume that orbifolding commutes with
part of the U-duality group.  Closely related issues have been studied
by Sen\cite{ashoke} who has constructed examples where the assumption
works as well as those for which the assumption fails. Our case seems
to be one where the assumption works.

\section{M(atrix) theory compactifications in seven dimensions}

In this section, we will first discuss the relevant details of
Heterotic M(atrix) theory on $T^3$ followed by M(atrix) theory on
$T^4/\BZ_2$. We then describe the conjecture and show that it implies
the seven dimensional duality between Heterotic string theory on $T^3$
and M-theory on $K3$. We shall assume that both theories should be
described by some 5+1 dimensional theory just as in the case of
$T^4$ compactification of M(atrix) theory\cite{rozali}.

The Heterotic string and its compactifications have been the target of
recent studies in M(atrix) theory\cite{het,horava}. The result which
we obtain from these papers is the following, M(atrix) theory on
$(S^1/\BZ_2) \times T^3$ is equivalent to the (orbifold) Yang-Mills
theory on $\widetilde{S}^1 \times (\widetilde{T}^3/\BZ_2)$.  Let us
assume that the tori are rectangular and the circles have radii $R_i$,
$i=1,\ldots,4$ with $i=1$ corresponding to the circle on which the
orbifold $\BZ_2$ acts. The base space for the Yang-Mills theory is
given by the dual torus obtained from circles of radii (in units where
the eleven dimensional Planck scale $l_{11}=1$ and we have ignored
some constant factors which do not affect our discussion)
\begin{equation}
\Sigma_i \sim 1/(r R_i)\quad,\label{edual}
\end{equation}
where $r$ is the radius of the eleventh dimension. The $\BZ_2$ group
acts by reflection on the coordinates along the $2,3,4$ directions of
the Yang-Mills group.  Without the orbifolding, we would be discussing
the case of M(atrix) theory on $T^4$. In this case, Rozali has shown
that an extra dimension of radius
\begin{equation}
\Sigma_5 \sim 1/(r R_1 R_2 R_3 R_4)
\end{equation}
emerges\footnote{This involves making the assumption 
that the extra dimension emerges even though there
are fewer supersymmetries here. This assumption works because
the supersymmetry is broken only by the boundary conditions and hence
away from the fixed points all the supersymmetries are seen. We thank
the referee for this remark.}.
The momentum modes around this dimension
correspond to zero size instantons with the instanton number being
identified with the momentum.  Since the $\BZ_2$ acts as a reflection
on odd number of dimensions, the instanton number changes sign under
the $\BZ_2$. The change in the sign of the instanton number
corresponds to a reflection in the 5th dimension. Thus we obtain the
complete action of $\BZ_2$ on the five coordinates (i.e., we are
including the extra dimension as a fifth coordinate) to be
\begin{equation}
\sigma^1 \rightarrow \sigma^1\quad{\rm and}\quad
\sigma^i \rightarrow - \sigma^i\quad{\rm for}\
i=2,\ldots,5\quad.\label{etra}
\end{equation}
This implies that heterotic M(atrix) theory on $T^3$ is described by
a 5+1 dimensional theory with coordinates $\sigma^i$, $i=1,\ldots,5$.
\footnote{We shall defer a discussion of this theory to the
conclusion.}  From the $\BZ_2$ action, we see that the last four
coordinates form a $\widetilde{K3}$ at its $\BZ_2$ orbifold limit.

We shall next consider M(atrix) compactification on $K3$ (in its
orbifold limit $T^4/\BZ_2$) with the $T^4$ being rectangular with
circles of radii $R_i^{\prime}$. (We shall indicate all parameters on
the $K3$ side by a prime. The unprimed parameters always refer to the
Heterotic side.)  The M(atrix) theory is obtained from the (orbifold)
Yang-Mills theory on the dual torus $\widetilde{T}^4/\BZ_2$.  The
radii of the circles of the dual torus are given by a formula similar
to eqn. (\ref{edual})
\begin{equation}
 \Sigma_i^{\prime}\sim 1/(r R_i^{\prime})\quad.
\end{equation}
Again, Rozali's argument provides us with an extra dimension of radius
\begin{equation}
\Sigma_5^{\prime} \sim 1/(r R_1^{\prime} R_2^{\prime} R_3^{\prime}
R_4^{\prime})\quad. 
\end{equation}

The action of the orbifold $\BZ_2$ on this extra dimension is again
obtained by considering its action on instanton number. Unlike the
heterotic case, there is no change in sign here and hence the $\BZ_2$
acts trivially on the fifth dimension. Explicitly, it has the
following action on the five coordinates (with the fifth coordinate
being the extra dimension)
\begin{equation}
\sigma^{\prime~i} \rightarrow - \sigma^{\prime~i}\quad{\rm for}
\quad i=1,\ldots,4\quad {\rm and}\quad
\sigma^{\prime~5} \rightarrow \sigma^{\prime~5}\quad.\label{etrb}
\end{equation}
Thus, we obtain that M(atrix) theory on $K3$ is described by a 5+1
dimensional theory with coordinates $\sigma'^i$, $i=1,\ldots,5$.  From
the $\BZ_2$ action, we see that the first four coordinates form a
$\widetilde{K3}$ at its $\BZ_2$ orbifold limit.

Comparing the action of the $\BZ_2$ in both the cases of interest as
given in eqns. (\ref{etra}) and (\ref{etrb}), we can map the two
$\BZ_2$'s into one another by exchanging $\sigma^1$ and
$\sigma^5$. From the viewpoint of Yang-Mills theory, this is a
non-trivial operation since the fifth dimension is not manifest.
However, this operation is an element of the U-duality group and if
U-duality commutes with orbifolding, this exchange is allowed. Thus,
this leads to the following conjecture:

\noindent
{\it Heterotic theory on $T^3$ and M(atrix) theory on $K3$ are
described by the RG fixed point of the same 5+1 dimensional theory
with base $\widetilde{S}^1\times \widetilde{K3}$.  The orbifold limit
of this theory can be obtained by orbifolding the 5+1 dimensional
theory with base $\widetilde{T}^5$ which describes M(atrix) theory on
$T^4$.}

Let us now check whether this conjecture can be true. The conjecture
provides a relationship between the heterotic and $K3$
theories.  Equating the radius of $\widetilde{S}^1$ and volume of the
$\widetilde{K3}$ in the 5+1 dimensional theory for both theories, we
get
\begin{eqnarray}
\Sigma_1 &\sim& \Sigma_5^{\prime}\quad, \\
\Sigma_2\Sigma_3\Sigma_4\Sigma_5 &\sim& 
\Sigma_1^{\prime}\Sigma_2^{\prime}\Sigma_3^{\prime}\Sigma_4^{\prime}\quad.
\end{eqnarray}
This leads to the following relationships in target spacetime:
\begin{eqnarray}
R_1 &\sim&  V_{K3}^{\prime}\quad,\\
V_{T^3} &\sim& 1\quad,
\end{eqnarray}
where $V_{T_3}=R_2 R_3 R_4$ and
$V_{K3}^{\prime}=R_1^{\prime}R_2^{\prime}R_3^{\prime}R_4^{\prime}$ are
volumes measured in units where $l_{11}=1$. The parameters of the
Heterotic string are
\begin{equation}
\alpha^{\prime}=R_1\quad{\rm and}\quad 1/\lambda_7^2 = V_{T^3}/R_1^{3/2}\quad,
\label{ehet}
\end{equation}
where $\alpha^{\prime}$ is the string tension and $\lambda_7$ is the
seven dimensional string coupling constant\footnote{The string tension
follows from observing that the heterotic string is obtained by
wrapping the M2-brane around the orbifold circle of radius $R_1$. The
string coupling is obtained from standard dimensional reduction of the
ten dimensional relation\cite{hw} $\lambda_{10}^2=R_1^3$ in units
where $l_{11}$ is unity. See \cite{witten,duff} for example.}.

Rewriting eqn. (\ref{ehet}) using (\ref{etra}) and (\ref{etrb}),  we obtain 
that 
\begin{equation}
\alpha^{\prime}\sim V_{K3}^{\prime}\quad {\rm and}\quad \lambda_7\sim
(V_{K3}^{\prime})^{3/4}
\label{erel}
\end{equation}
which is in agreement with seven dimensional duality\cite{witten}. The
string tension being proportional to the volume of $K3$ agrees with
the fact that the heterotic string arises from wrapping an M5-brane on
the $K3$. Also, $V_{T^3}\sim1$ shows that in making the correspondence
one has to hold the volume of the three torus to be a constant
independent of $V_{K3}$.  We would like to emphasise that the
relationships obtained in eqn. (\ref{erel}) are based on the
conjecture and hence distinct from (for example) the derivation of
Heterotic string tension in ref.  \cite{kthree} for M(atrix) theory on
K3.

\section{Conclusion}

In this note, we have provided evidence for the conjecture that both
seven dimensional theories with eight supercharges are described by
the same 5+1 dimensional field theory. The assumption that orbifolding
commutes with U-duality seems to be reasonable in this case.  For
M(atrix) compactification on $K3$ (away from the orbifold limit), the
5+1 dimensional theory has as its base space $\widetilde{K3}\times
\widetilde{S}^1$ and has ${\cal N}=1$ supersymmetry. It would be of
interest to work out the precise relationship between the targetspace
$K3$ and base $\widetilde{K3}$ for generic (non-orbifold) K3.

What can we say with regard to the 5+1 dimensional theory?  In
M-theory on $K3$, the heterotic string is obtained by wrapping the
M5-brane around the $K3$. This suggests that the $5+1$ dimensional
theory should be related in some sense to a theory of N coincident
M5-branes with worldvolume $K3\times S^1\times \BR$. The world volume
field theory of an M5-brane is described by a chiral ${\cal N}=2$ (16
supersymmetries) with a single tensor multiplet. N coincident
M5-branes will have N tensor multiplets, with extra tensor multiplets
in the adjoint of U(N) describing M2-branes connecting the
M5-branes\cite{fivebrane}. The wrapping of the M5-branes on $K3$
breaks half the supersymmetry on the worldvolume leading to ${\cal
N}=1$ supersymmetry as required. However at present there does not seem to
an understanding of the theory described above. It is thus better
to view the theory as the one given by compactifying the M(atrix) theory 
describing M-theory on $T^4$, the theory proposed in ref. \cite{brs},
on $S^1 \times $K3.

\noindent {\it Note added}:
The results reported here have also been independently derived
by P. Ho\v rava\cite{dual}, S.-J. Rey\cite{rey} and  Berkooz and 
Rozali\cite{bz}.

\acknowledgments
This work is supported by a
fellowship from the Alexander von Humboldt Foundation. I would like to
thank P. Ho\v rava and T. Jayaraman for useful conversations.  
In addition, I would
like to thank Werner Nahm and the theory group at Bonn for their
hospitality. We are grateful to  the referee for correcting some erroneous 
as well as misleading statements which has improved the quality of this paper.

\end{document}